\documentstyle[10pt,epsf,epsfig,dp_delphititle,oldlfont]{dp_delphi}
\makeindex
\def\DpPaperGroup{EP}
\def\DpPaperRef{2001-033 }
\def\DpDate{03 May 2001}
\def\DpAuthors{DELPHI Collaboration}
\def\DpSubmit{(Accepted by Phys.Lett.B)}
\def\DpTitle{{Single Intermediate Vector Boson Production 
     in \mbox{\boldmath $e^+ e^-$} collisions at 
     \mbox{\boldmath $\sqrt{s} = 183 $} and \mbox{\boldmath $189$} GeV}}
\def\DpComment{ }
\def\DpEMail{ }
\newcommand{\be}{\begin{equation}}
\newcommand{\ee}{\end{equation}}
\newcommand{\bt}{\begin{itemize}}
\newcommand{\et}{\end{itemize}}
\newcommand{\ben}{\begin{enumerate}}
\newcommand{\een}{\end{enumerate}}
\def\lu{{\cal {L}}_{int}}
\def\pbi{pb$^{-1}$}
\def\epem{e^+e^-}
\def\qq{q\bar{q}}
\def\mm{\mu^+ \mu^-}

\def\eell{e^+e^-\to l^+l^-}

\def\eeww{e^+e^-\to W^+W^-}

\def\eegz{e^+e^-\to Z\gamma}
\def\eewe{e^+e^-\to e^- \bar{\nu}_e W^+}
\def\eeze{e^+e^-\to e^+e^-Z}
\def\qqtv{q\bar{q}'\tau\bar{\nu}_{\tau}}
\def\wev{e\nu_eW}
\def\evqq{e^-\bar{\nu}_e{q}\bar{q}'}
\def\evlv{e^-\bar{\nu}_el^+\nu_l}
\def\evmv{e^-\bar{\nu}_e\mu^+\nu_\mu}
\def\evtv{e^-\bar{\nu}_e\tau^+\nu_\tau}
\def\evev{e\nu e\nu}
\def\gs{\gamma^{*}}
\def\deg{^{\circ}}
\def\qbar{\bar{q}}
\def\fbar{\bar{f}}
\begin{document}
%%%%%%%%%%%%%%%%%%%%%%%%%% They are a problem with Coll.Sty ?
\makeatletter
%\input{dp_system:coll.sty}
% Collapse citation numbers to ranges.  Non-numeric and undefined labels
% are handled.  No sorting is done.  E.g., 1,3,2,3,4,5,foo,1,2,3,?,4,5
% gives 1,3,2-5,foo,1-3,?,4,5
\newcount\@tempcntc
\def\@citex[#1]#2{\if@filesw\immediate\write\@auxout{\string\citation{#2}}\fi
  \@tempcnta\z@\@tempcntb\m@ne\def\@citea{}\@cite{\@for\@citeb:=#2\do
    {\@ifundefined
       {b@\@citeb}{\@citeo\@tempcntb\m@ne\@citea\def\@citea{,}{\bf ?}\@warning
       {Citation `\@citeb' on page \thepage \space undefined}}%
    {\setbox\z@\hbox{\global\@tempcntc0\csname b@\@citeb\endcsname\relax}%
     \ifnum\@tempcntc=\z@ \@citeo\@tempcntb\m@ne
       \@citea\def\@citea{,}\hbox{\csname b@\@citeb\endcsname}%
     \else
      \advance\@tempcntb\@ne
      \ifnum\@tempcntb=\@tempcntc
      \else\advance\@tempcntb\m@ne\@citeo
      \@tempcnta\@tempcntc\@tempcntb\@tempcntc\fi\fi}}\@citeo}{#1}}
\def\@citeo{\ifnum\@tempcnta>\@tempcntb\else\@citea\def\@citea{,}%
  \ifnum\@tempcnta=\@tempcntb\the\@tempcnta\else
   {\advance\@tempcnta\@ne\ifnum\@tempcnta=\@tempcntb \else \def\@citea{--}\fi
    \advance\@tempcnta\m@ne\the\@tempcnta\@citea\the\@tempcntb}\fi\fi}
 
\makeatother
%%%%%%%%%%%%%%%%%%%%%%%%%% ??????????????????????????????????
% Generate the title page
\begin{titlepage}
\pagenumbering{roman}
\CERNpreprint{\DpPaperGroup}{\DpPaperRef} % Reference of the paper
\date{{\small\DpDate}} % Date of the paper
\title{\DpTitle} % Title of the paper
\address{\DpAuthors} % General name of the author(s)
\begin{shortabs} % Start the abstract
\noindent
%   abstract.tex
%
\noindent
%===================> Abstract     =====> To be filled <=====%
The cross-sections for the production of single charged and neutral
intermediate vector bosons were measured using integrated luminosities of 
52~pb$^{-1}$ and 154~pb$^{-1}$ collected by the DELPHI experiment at 
centre-of-mass energies of 182.6~GeV and 188.6~GeV, respectively. 
The cross-sections for the reactions were determined in limited
kinematic regions. The results found are in agreement with the Standard
Model predictions for these channels.
\end{shortabs}
\vfill
\begin{center}
\DpSubmit \ \\ % Horrible hack to allow to have DpSubmit empty
\DpComment \ \\
\DpEMail \ \\
\end{center}
\vfill
\clearpage
\headsep 10.0pt
\addtolength{\textheight}{10mm}
\addtolength{\footskip}{-5mm}
\begingroup
% Commands to process the author names
%
\newcommand{\DpName}[2]{\hbox{#1$^{\ref{#2}}$},\hfill}
\newcommand{\DpNameTwo}[3]{\hbox{#1$^{\ref{#2},\ref{#3}}$},\hfill}
\newcommand{\DpNameThree}[4]{\hbox{#1$^{\ref{#2},\ref{#3},\ref{#4}}$},\hfill}
\newskip\Bigfill \Bigfill = 0pt plus 1000fill
\newcommand{\DpNameLast}[2]{\hbox{#1$^{\ref{#2}}$}\hspace{\Bigfill}}
%
%\small
\footnotesize
\noindent
\DpName{P.Abreu}{LIP}
\DpName{W.Adam}{VIENNA}
\DpName{T.Adye}{RAL}
\DpName{P.Adzic}{DEMOKRITOS}
\DpName{I.Ajinenko}{SERPUKHOV}
\DpName{Z.Albrecht}{KARLSRUHE}
\DpName{T.Alderweireld}{AIM}
\DpName{G.D.Alekseev}{JINR}
\DpName{R.Alemany}{CERN}
\DpName{T.Allmendinger}{KARLSRUHE}
% ...
\normalsize
\endgroup
\titlefoot{Department of Physics and Astronomy, Iowa State
     University, Ames IA 50011-3160, USA
    \label{AMES}}
\titlefoot{Physics Department, Univ. Instelling Antwerpen,
     Universiteitsplein 1, B-2610 Antwerpen, Belgium \\
     \indent~~and IIHE, ULB-VUB,
     Pleinlaan 2, B-1050 Brussels, Belgium \\
     \indent~~and Facult\'e des Sciences,
     Univ. de l'Etat Mons, Av. Maistriau 19, B-7000 Mons, Belgium
    \label{AIM}}
\titlefoot{Physics Laboratory, University of Athens, Solonos Str.
     104, GR-10680 Athens, Greece
    \label{ATHENS}}
\titlefoot{Department of Physics, University of Bergen,
     All\'egaten 55, NO-5007 Bergen, Norway
    \label{BERGEN}}
\titlefoot{Dipartimento di Fisica, Universit\`a di Bologna and INFN,
     Via Irnerio 46, IT-40126 Bologna, Italy
    \label{BOLOGNA}}
\titlefoot{Centro Brasileiro de Pesquisas F\'{\i}sicas, rua Xavier Sigaud 150,
     BR-22290 Rio de Janeiro, Brazil \\
     \indent~~and Depto. de F\'{\i}sica, Pont. Univ. Cat\'olica,
     C.P. 38071 BR-22453 Rio de Janeiro, Brazil \\
     \indent~~and Inst. de F\'{\i}sica, Univ. Estadual do Rio de Janeiro,
     rua S\~{a}o Francisco Xavier 524, Rio de Janeiro, Brazil
    \label{BRASIL}}
\titlefoot{Comenius University, Faculty of Mathematics and Physics,
     Mlynska Dolina, SK-84215 Bratislava, Slovakia
    \label{BRATISLAVA}}
\titlefoot{Coll\`ege de France, Lab. de Physique Corpusculaire, IN2P3-CNRS,
     FR-75231 Paris Cedex 05, France
    \label{CDF}}
\titlefoot{CERN, CH-1211 Geneva 23, Switzerland
    \label{CERN}}
\titlefoot{Institut de Recherches Subatomiques, IN2P3 - CNRS/ULP - BP20,
     FR-67037 Strasbourg Cedex, France
    \label{CRN}}
\titlefoot{Now at DESY-Zeuthen, Platanenallee 6, D-15735 Zeuthen, Germany
    \label{DESY}}
\titlefoot{Institute of Nuclear Physics, N.C.S.R. Demokritos,
     P.O. Box 60228, GR-15310 Athens, Greece
    \label{DEMOKRITOS}}
\titlefoot{FZU, Inst. of Phys. of the C.A.S. High Energy Physics Division,
     Na Slovance 2, CZ-180 40, Praha 8, Czech Republic
    \label{FZU}}
\titlefoot{Dipartimento di Fisica, Universit\`a di Genova and INFN,
     Via Dodecaneso 33, IT-16146 Genova, Italy
    \label{GENOVA}}
\titlefoot{Institut des Sciences Nucl\'eaires, IN2P3-CNRS, Universit\'e
     de Grenoble 1, FR-38026 Grenoble Cedex, France
    \label{GRENOBLE}}
\titlefoot{Helsinki Institute of Physics, HIP,
     P.O. Box 9, FI-00014 Helsinki, Finland
    \label{HELSINKI}}
\titlefoot{Joint Institute for Nuclear Research, Dubna, Head Post
     Office, P.O. Box 79, RU-101 000 Moscow, Russian Federation
    \label{JINR}}
\titlefoot{Institut f\"ur Experimentelle Kernphysik,
     Universit\"at Karlsruhe, Postfach 6980, DE-76128 Karlsruhe,
     Germany
    \label{KARLSRUHE}}
\titlefoot{Institute of Nuclear Physics and University of Mining and Metalurgy,
     Ul. Kawiory 26a, PL-30055 Krakow, Poland
    \label{KRAKOW}}
\titlefoot{Universit\'e de Paris-Sud, Lab. de l'Acc\'el\'erateur
     Lin\'eaire, IN2P3-CNRS, B\^{a}t. 200, FR-91405 Orsay Cedex, France
    \label{LAL}}
\titlefoot{School of Physics and Chemistry, University of Lancaster,
     Lancaster LA1 4YB, UK
    \label{LANCASTER}}
\titlefoot{LIP, IST, FCUL - Av. Elias Garcia, 14-$1^{o}$,
     PT-1000 Lisboa Codex, Portugal
    \label{LIP}}
\titlefoot{Department of Physics, University of Liverpool, P.O.
     Box 147, Liverpool L69 3BX, UK
    \label{LIVERPOOL}}
\titlefoot{LPNHE, IN2P3-CNRS, Univ.~Paris VI et VII, Tour 33 (RdC),
     4 place Jussieu, FR-75252 Paris Cedex 05, France
    \label{LPNHE}}
\titlefoot{Department of Physics, University of Lund,
     S\"olvegatan 14, SE-223 63 Lund, Sweden
    \label{LUND}}
\titlefoot{Universit\'e Claude Bernard de Lyon, IPNL, IN2P3-CNRS,
     FR-69622 Villeurbanne Cedex, France
    \label{LYON}}
\titlefoot{Univ. d'Aix - Marseille II - CPP, IN2P3-CNRS,
     FR-13288 Marseille Cedex 09, France
    \label{MARSEILLE}}
\titlefoot{Dipartimento di Fisica, Universit\`a di Milano and INFN-MILANO,
     Via Celoria 16, IT-20133 Milan, Italy
    \label{MILANO}}
\titlefoot{Dipartimento di Fisica, Univ. di Milano-Bicocca and
     INFN-MILANO, Piazza delle Scienze 2, IT-20126 Milan, Italy
    \label{MILANO2}}
\titlefoot{IPNP of MFF, Charles Univ., Areal MFF,
     V Holesovickach 2, CZ-180 00, Praha 8, Czech Republic
    \label{NC}}
\titlefoot{NIKHEF, Postbus 41882, NL-1009 DB
     Amsterdam, The Netherlands
    \label{NIKHEF}}
\titlefoot{National Technical University, Physics Department,
     Zografou Campus, GR-15773 Athens, Greece
    \label{NTU-ATHENS}}
\titlefoot{Physics Department, University of Oslo, Blindern,
     NO-1000 Oslo 3, Norway
    \label{OSLO}}
\titlefoot{Dpto. Fisica, Univ. Oviedo, Avda. Calvo Sotelo
     s/n, ES-33007 Oviedo, Spain
    \label{OVIEDO}}
\titlefoot{Department of Physics, University of Oxford,
     Keble Road, Oxford OX1 3RH, UK
    \label{OXFORD}}
\titlefoot{Dipartimento di Fisica, Universit\`a di Padova and
     INFN, Via Marzolo 8, IT-35131 Padua, Italy
    \label{PADOVA}}
\titlefoot{Rutherford Appleton Laboratory, Chilton, Didcot
     OX11 OQX, UK
    \label{RAL}}
\titlefoot{Dipartimento di Fisica, Universit\`a di Roma II and
     INFN, Tor Vergata, IT-00173 Rome, Italy
    \label{ROMA2}}
\titlefoot{Dipartimento di Fisica, Universit\`a di Roma III and
     INFN, Via della Vasca Navale 84, IT-00146 Rome, Italy
    \label{ROMA3}}
\titlefoot{DAPNIA/Service de Physique des Particules,
     CEA-Saclay, FR-91191 Gif-sur-Yvette Cedex, France
    \label{SACLAY}}
\titlefoot{Instituto de Fisica de Cantabria (CSIC-UC), Avda.
     los Castros s/n, ES-39006 Santander, Spain
    \label{SANTANDER}}
\titlefoot{Dipartimento di Fisica, Universit\`a degli Studi di Roma
     La Sapienza, Piazzale Aldo Moro 2, IT-00185 Rome, Italy
    \label{SAPIENZA}}
\titlefoot{Inst. for High Energy Physics, Serpukov
     P.O. Box 35, Protvino, (Moscow Region), Russian Federation
    \label{SERPUKHOV}}
\titlefoot{J. Stefan Institute, Jamova 39, SI-1000 Ljubljana, Slovenia
     and Laboratory for Astroparticle Physics,\\
     \indent~~Nova Gorica Polytechnic, Kostanjeviska 16a, SI-5000 Nova Gorica, Slovenia, \\
     \indent~~and Department of Physics, University of Ljubljana,
     SI-1000 Ljubljana, Slovenia
    \label{SLOVENIJA}}
\titlefoot{Fysikum, Stockholm University,
     Box 6730, SE-113 85 Stockholm, Sweden
    \label{STOCKHOLM}}
\titlefoot{Dipartimento di Fisica Sperimentale, Universit\`a di
     Torino and INFN, Via P. Giuria 1, IT-10125 Turin, Italy
    \label{TORINO}}
\titlefoot{Dipartimento di Fisica, Universit\`a di Trieste and
     INFN, Via A. Valerio 2, IT-34127 Trieste, Italy \\
     \indent~~and Istituto di Fisica, Universit\`a di Udine,
     IT-33100 Udine, Italy
    \label{TU}}
\titlefoot{Univ. Federal do Rio de Janeiro, C.P. 68528
     Cidade Univ., Ilha do Fund\~ao
     BR-21945-970 Rio de Janeiro, Brazil
    \label{UFRJ}}
\titlefoot{Department of Radiation Sciences, University of
     Uppsala, P.O. Box 535, SE-751 21 Uppsala, Sweden
    \label{UPPSALA}}
\titlefoot{IFIC, Valencia-CSIC, and D.F.A.M.N., U. de Valencia,
     Avda. Dr. Moliner 50, ES-46100 Burjassot (Valencia), Spain
    \label{VALENCIA}}
\titlefoot{Institut f\"ur Hochenergiephysik, \"Osterr. Akad.
     d. Wissensch., Nikolsdorfergasse 18, AT-1050 Vienna, Austria
    \label{VIENNA}}
\titlefoot{Inst. Nuclear Studies and University of Warsaw, Ul.
     Hoza 69, PL-00681 Warsaw, Poland
    \label{WARSZAWA}}
\titlefoot{Fachbereich Physik, University of Wuppertal, Postfach
     100 127, DE-42097 Wuppertal, Germany
    \label{WUPPERTAL}}
\addtolength{\textheight}{-10mm}
\addtolength{\footskip}{5mm}
\clearpage
\headsep 30.0pt
\end{titlepage}
%%%%%%%%%%%%%%%%%%%%%%%%%
%
% Change for the document body
%%\pagestyle{heading} % for page numbering
\pagenumbering{arabic} % page numbering in number
\setcounter{footnote}{0} %
\large
%\linenumbers %%%CD
\section{Introduction}

The measurement of the production cross-section of a single 
vector boson ($\eewe$, $\eeze$)\footnote{Charge conjugate states are
  implied throughout the text.} is a test of the Standard Model.  
In addition, the study of these processes is important in  
the evaluation of background to the search for the Higgs boson 
and for physics beyond the Standard Model.
Single-$W$ production is also interesting in itself for the measurement
of the  trilinear couplings at the $WW\gamma$ vertex; 
this measurement, in combination with other physics channels, has been
made by the {\mbox DELPHI} Collaboration  and is reported
elsewhere~\cite{paptgc}. 

This letter presents measurements of single-$W$ and single-$Z$
production cross-sections using the data collected by {\mbox DELPHI}
at centre-of-mass energies of 182.6 and 188.6~GeV.  
The corresponding integrated luminosities are 52 and 154~\pbi, respectively.

The criteria for the selection of the events are mainly based on the 
information from the tracking system, the calorimeters and the muon chambers
of the {\mbox DELPHI} detector. 
A detailed description of the {\mbox DELPHI} apparatus and its
performances can be
found in~\cite{DELPHI,DELPHIperf}. The detector has remained
essentially unchanged in the past few years, except for upgrades of the 
Vertex Detector~\cite{VD}.

\section{Definition of the signal}
%\footnote{The variables used for the signal definition in this section were considered at the generator level.}}

Single boson production is investigated in this paper through
four-fermions final states, $e^- \bar{\nu}_e f\bar{f}'$ and $e^+ e^- f\bar{f}$. 
%Diagrams with different topologies, for example conversion
%diagrams for $e\nu f\bar{f}'$ final state and single and double
%resonant diagrams in the case of $e^+ e^- f\bar{f}$ final
%states, contribute to the cross-sections as well~\cite{YRlep2}. 
These final states receive contributions from single resonant diagrams
producing, respectively, the $W$ and $Z$  signals studied here, and
from other diagrams, including doubly resonant production, conversion
diagrams and multiperipheral processes~\cite{YRlep2}.
To enhance the single boson production contribution, the
cross-sections correspond to the limited kinematic regions described below.
The measured cross-sections therefore refer to the entire set of
diagrams contributing to the specific final states, with the exception
of multiperipheral diagrams~\cite{YRlep2} whose contributions were
evaluated separately and then subtracted.

\paragraph{\mbox{\boldmath $\wev$} {\bf channel:} }
The four-fermion final states $e^- \bar{\nu}_e q \bar{q}'$ and $e^-
\bar{\nu}_e l^+ \nu_l$   ($l=\mu,\tau$) can be produced both via
single-$W$ production, referred to
as $\wev$ in the following, or via $W$-pair production.
A distinctive feature of $\wev$ is the fact that the distribution of the
electron direction is strongly peaked at small polar angles with 
respect to the incoming electron beam direction. 
The signal definition was restricted to the region of phase space where the
contribution of the single-$W$ process is dominant.
The polar angle of the outgoing electron, $\theta_{e^-}$, was required to be smaller than
the lower edge of the {\mbox DELPHI} detector acceptance:
\begin{eqnarray} 
|\cos\theta_{e^-}| > 0.9993.
\label{eq:cut1}
\end{eqnarray} 

\noindent
Additional selections were applied to avoid the phase space
regions of low $f\bar{f}'$ invariant mass, mostly due to multiperipheral
diagrams, where large uncertainties affect the cross-section
computation. It was required that:
\begin{eqnarray} 
m_{q\bar{q}'} > 45 \ \mbox{GeV}/c^2 \ \ &
\mbox{for} & \ \ e^{-}\bar{\nu}_{e} q \bar{q}', \\
E_{l^+} > 15 \ \mbox{GeV} \ \  &
\mbox{for} & \ \ e^{-}\bar{\nu}_{e} l^{+}\nu_{l}
 \ \ (l^{+}= \mu^{+}, \tau^{+}), \nonumber 
\label{eq:cut2}
\end{eqnarray}
where $m_{q\bar{q}'}$ is the $q\bar{q}'$ invariant mass and $E_{l^+}$
the lepton energy.

Single-$W$ production accounts for more than $90\%$ of all 
$e^{-}\bar{\nu}_{e} q \bar{q}'$ and $e^{-}\bar{\nu}_{e} l^{+}\nu_{l}$ 
events in the kinematic region defined above.
The sum of the cross-sections in the channels $e^-\bar{\nu}_e q\bar{q}'$,
$e^-\bar{\nu}_e \mu^+\nu_{\mu}$, $e^-\bar{\nu}_e \tau^+\nu_{\tau}$, hereafter
called $e\nu f\bar{f}'$, is then compared to the 
theoretical calculations from \mbox{GRC4F}~\cite{GRC4F} and
\mbox{WPHACT}~\cite{WPHACT}. 

The $e^-\bar{\nu}_e e^+\nu_e$ contributions are treated separately
(below) because they have $t$-channel
contributions both for single-$W$ and single-$Z$ topologies which are not
easily disentangled.

\paragraph{\mbox{\boldmath $e\nu e\nu$} {\bf channel:}}
In the kinematic region defined above, this final state receives, 
besides single-$W$ production, a large contribution from
$Ze^+e^-$ production (with $Z \to \nu_e\bar{\nu}_e$) and from the 
interference between single-$W$ and $Z\epem$ processes. 
In addition it is not possible, experimentally, to disentangle the 
$e^{+} \nu_{e} e^{-} \bar{\nu}_e$ final state from the 
$e^{+} e^{-}  \nu_{\mu} \bar{\nu}_{\mu}$ and 
$e^{+} e^{-} \nu_{\tau} \bar{\nu}_{\tau}$ final states with the
neutrinos produced
in $Z \to \nu_{\mu}\bar{\nu}_{\mu}, \nu_{\tau}\bar{\nu}_{\tau}$
decays. 
Therefore, a topological cross-section was defined corresponding to the 
$e\nu e\nu$ final state (where a sum over all neutrino flavours is implied), 
to be compared with theoretical calculations. 
This was done restricting further the signal phase space to:

\begin{equation}
|\cos\theta_{e^+}| < 0.72, ~~~~
E_{e^+} > 30 ~\mbox{GeV}.  
\label{eq:cut3}
\end{equation}

\paragraph{\mbox{\boldmath $\gs/Zee$} {\bf channel:}}
The neutral bosons are produced in the so-called electroweak Compton scattering
process $e \gamma \to  e \gs /Z$, where a quasi-real photon is radiated from 
one of the beam electrons and scattered off the other beam~\cite{singleZ}. 
In this paper only decays of the $\gs/Z$ into hadrons and $\mm$ pairs 
have been considered:
the signature of such events is an electron, typically of low energy,
recoiling against the $\gs/Z$ system, with the other electron
usually lost in the beam-pipe. 
The signal cross-section presented in this note refers to the overall set of
graphs contributing to the $e^+e^-f\fbar$ $(f=q,\mu)$ final state with
the exception of the so-called multiperipheral ones, typical of the 
$\gamma \gamma$ physics~\cite{YRlep2}. 
The signal was defined topologically, requiring
at least one electron (tag electron) to be within the acceptance of the
{\mbox DELPHI} forward electromagnetic calorimeter, i.e. $|\cos \theta_{e^+}|
<0.985$, and having an energy $E_{e^+} > 4$~GeV.
The measured cross-section was then compared with that obtained from the
\mbox{GRC4F} program. A selection on the minimum energy of this visible electron,  
$E_{e^+} > 1$~GeV, was used in the computation of the cross-section to
avoid numerical instabilities in the integration.

\paragraph{} For both the $\wev$ and $Zee$ samples,
signal events were simulated with the \mbox{GRC4F} event generator. 
For background processes, different
generators were used: \mbox{EXCALIBUR}~\cite{excal} for the $WW$ and other
four-fermion final states, \mbox{PYTHIA}~\cite{lund} for
$q\qbar(\gamma)$, \mbox{TEEGG}~\cite{teeg} and BHWIDE~\cite{bhwide} for
$e^+e^-\to{e^+}e^-\gamma$, 
\mbox{KORALZ}~\cite{KORALZ} for $e^+e^-\to\mu^+\mu^-(\gamma),
\tau^+\tau^-(\gamma)$, \mbox{TWOGAM}~\cite{TWOGAM} 
and BDK~\cite{bdk} for two-photon collisions. 
All the events were processed through the full
{\mbox DELPHI} detector simulation and analysis chain~\cite{DELPHIperf}.

\section{Single-\mbox{\boldmath $W$} analysis}

Both the hadronic and the leptonic final states were considered in
the single-$W$ analysis. They are characterized
by the presence of two hadronic jets acoplanar with the beam or by
a single lepton with large transverse momentum, 
%LASTCORDB (begin)
respectively~\cite{alephl3}.
%LASTCORDB (end)

\subsection{Selection of hadronic events}

The experimental signature of $\evqq$ events consists of a pair
of acoplanar jets. The
undetected neutrino results in a large missing momentum at large angle
to the beam direction.

Other physics processes which can give rise to a similar topology are
$\eegz$ with $Z\to\qq$, $WW$ events with at least one $W$ decaying 
into hadrons, other four-fermion final states ($l^+l^-\qq$, 
$\nu\bar{\nu}\qq$, the latter being topologically identical to the signal) 
and two-photon collisions.
Some of these processes have cross-sections larger than that of the
signal by several orders of magnitude. Sequential cuts on the event
variables have been applied to reject them.

A sample of hadronic events was preselected by requiring at least seven
charged particles to be measured in the detector. The contribution from
two-photon collisions was reduced by requiring the opening angle of
the cone around the beam axis which contains 15\% of the visible
energy to be larger than 18$^\circ$: $\gamma\gamma$ events are
concentrated in the forward regions and have low values of this
variable. Furthermore, the total transverse momentum was required to be
larger than 16\% of $\sqrt{s}$.

The background from $\epem\to \qq(\gamma)$ was rejected by
requiring the effective collision energy, $\sqrt{s'}$~\cite{sprime}, to be
smaller than $0.85 \sqrt{s}$ and the cosine of the polar angle of the missing
momentum to satisfy the condition $|\cos\theta_{miss}|<0.9$. 
In addition, 
since the background is concentrated simultaneously 
at large $|\cos\theta_{miss}|$
and at values of $\sqrt{s'}$ close to the $Z$ mass, 
a selection on the correlation of the two variables was applied: 
$\sqrt{s'}>160\cdot|\cos\theta_{miss}|-30$~GeV. 
$Z(\gamma)$ events in which the ISR photon escaped
undetected in the dead region between the barrel and end-cap
electromagnetic calorimeters ($\theta\sim 40^\circ$) were suppressed
by looking for signals in the hermeticity counters
in a cone of 30$^\circ$ around the direction of the missing momentum.
%LASTCORDB (begin)

%LASTCORDB (end)
In addition,
a combined selection in the transverse momentum versus visible energy plane
was also applied to separate the $\evqq$ signal from the $WW$, $\nu\bar{\nu}\qq$ 
and $Z \gamma$ backgrounds.
Finally, the planarity of the three-body final state $\qq\gamma$ was
exploited: two jets were reconstructed with all the detected
particles, and their 
%LASTCORDB (begin)
acoplanarity\footnote{The acoplanarity was defined as the complementary to 180$^\circ$
 of the angle between the projections of the two jet directions in
 the plane transverse to the beam axis.} 
%LASTCORDB (end)
was required to exceed 15$^\circ$. 

\begin{figure}[htb]
\begin{center}
\epsfig{file=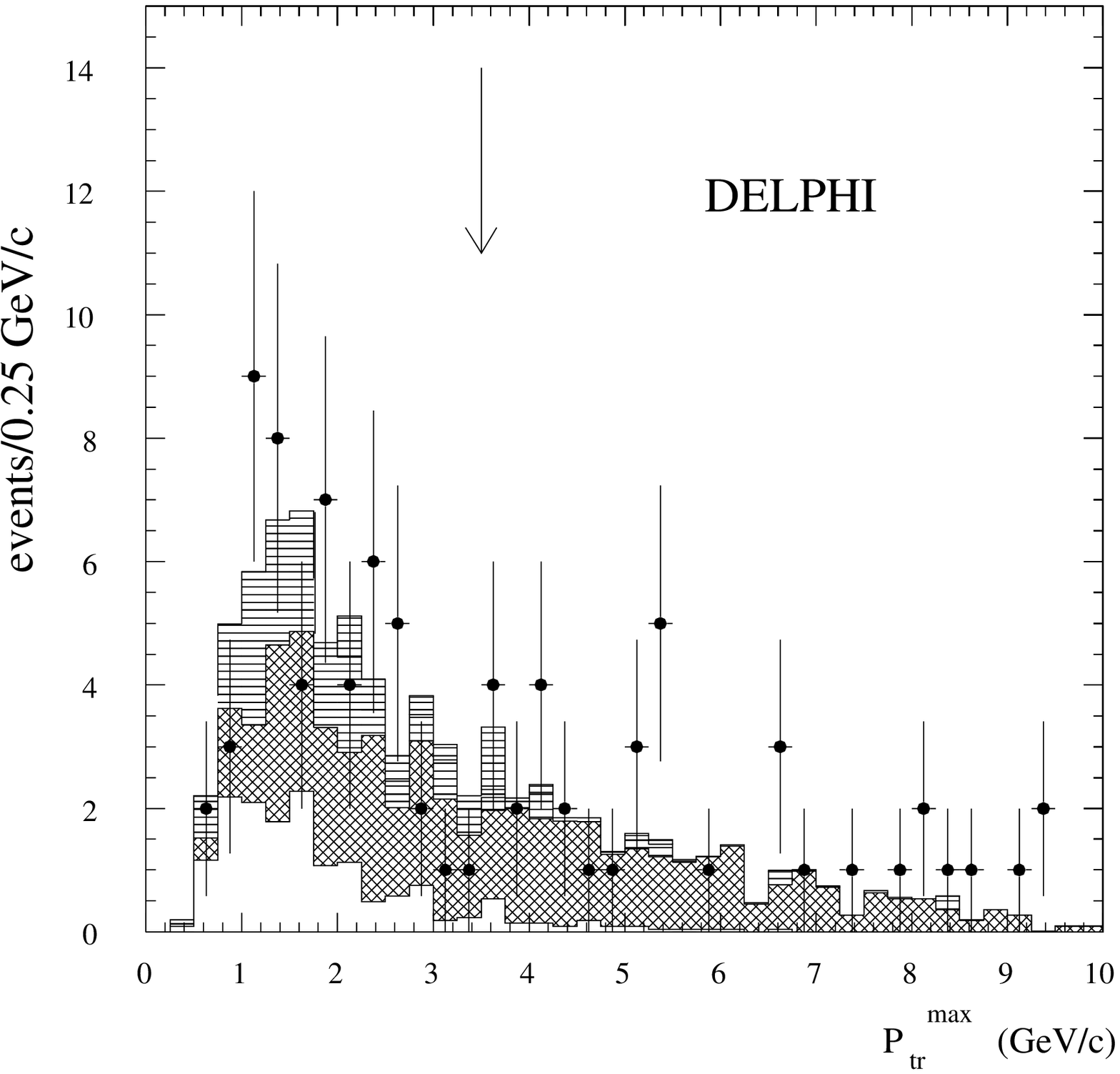,width=13cm}
\end{center}
\caption{$\wev$ channel $(W \to q\bar{q}')$ at $\sqrt{s}=189$~GeV. 
  Maximum transverse momentum of any particle with respect to the closest jet. 
  The open histogram represents the simulated $e^-\bar{\nu}_e q\bar{q}'$
  signal, the cross-hatched area represents the $WW$ background, the
  other backgrounds are shown with single hatching. Data points are
  indicated with statistical error bars. All the other selections
  described in Section~3.1 have already been applied. The arrow indicates
  the position of the cut on this variable.}
\label{fig:ptmax}
\end{figure}

The most important remaining contribution to the background is from
$W$-pair production. 
Events in which both $W$ bosons decay into a $q\bar{q}'$ pair tend
to have a four-jet topology, and were rejected by requiring the
distance parameter for the transition from 3 to 4 jets in the Durham
algorithm~\cite{durh}, $D^{join}_{3\to{4}}$, to be larger than 
0.005.  When one $W$ decays
into $q\bar{q}'$ and the other one into $l\bar{\nu}_{\ell}$, an
isolated lepton with high
energy is usually visible: events were rejected if an identified electron or
muon was found with an energy larger than 15~GeV and forming an angle
of more than 10$^\circ$ with the nearest track. If the lepton is a
$\tau$, the topology can be 3-jet-like: the $D^{join}_{2\to{3}}$ for the
transition from 2 to 3 jets was required to exceed 0.05. The
residual contamination from $\qqtv$ events in which the $\tau$ decay
products have very low energy or are very close to one of the hadronic
jets was reduced by a selection on the maximum transverse momentum of any
particle with respect to the closest jet, $P_{tr}^{max}<3.5$~GeV/$c$, as shown in Figure~\ref{fig:ptmax}.

\begin{table}[tb]
\begin{center}
\vspace{0.5cm}
\begin{tabular}{|l|c|c|c|c|c|c|r|}
\hline\hline
       &  $\wev$  &  $WW$  & $Z\gamma$ & $\nu\bar{\nu}\qq$ &
       Others & Total MC & Data
\\ \hline\hline
Step 1.
       & 36.3 & 1392.7 & 9484.2 & 29.5 & 263.8 & 11206.5 & 11550
\\ \hline
Step 2.
       & 22.6 &  565.1 &  503.5 & 17.3 & ~18.7 & ~1127.2 & 1130 
\\ \hline
Step 3.
       & 21.6 &  146.8 &  239.0 & 16.3 & ~15.5 & ~~439.2 & 405     
\\ \hline
Step 4.
       & $14.1\pm 0.7$ & $21.7\pm 0.9$ & $5.1 \pm 0.3$ & 
          $8.9\pm 0.2$ & $0.5 \pm 0.5$ & $50.3 \pm 1.3$ & 52 
\\ \hline\hline
\end{tabular}
\end{center}
\caption{Number of events expected from the contribution of different
       channels and observed in the data at different stages of the
       $\wev$, $W\to q\bar{q}'$ selection at $\sqrt{s} = 189$~GeV. The column
       labelled ``Others'' includes $l^+l^-\qq$ final states and
       two-photon collisions. Step 1 = after hadronic
       preselection and anti-$\gamma\gamma$ cuts; Step 2 = after cuts
       on $\sqrt{s'}$, $|\cos\theta_{miss}|$ and signal in the
       hermeticity counters; 
%LASTCORDB (begin)
%LASTCORDB       Step 3 = after all the cuts against $\qq(\gamma)$ events; 
       Step 3 = after the cuts on transverse momentum versus visible energy and on acoplanarity;
%LASTCORDB (end)
       Step 4 = the final sample after $WW$ rejection. 
       Details on the selection are provided in Section~3.1.}
\label{tab:nevts}
\end{table}

Table~\ref{tab:nevts} shows the number of selected events in the data at 189~GeV
in comparison to the expectation from the Monte Carlo simulation at successive
stages of the analysis. As can be seen from Table~\ref{tab:nevts}, the main
contamination in the final selected sample is due to $WW$ production,
with one $W$ decaying into hadrons and the other one into
$\tau\bar{\nu}_{\tau}$.

The efficiency of the selection for the signal, the expected background, 
the luminosity 
and the number of selected events in the data at the two centre-of-mass
energies are reported in Table~\ref{tab:sigqq}, together with the evaluated 
cross-section for the hadronic channel alone.

\begin{table}[tb]
\begin{center}
\vspace{0.5cm}
\begin{tabular}{|l|c|c|c|c||c|}
\hline\hline
$\sqrt{s}$ (GeV) &  Efficiency (\%) & $\sigma_{bgd}$ (pb)  & $\lu$ (pb$^{-1}$) &   $N_{data}$  &   $\sigma_{\evqq}$ (pb) \\
\hline \hline
 182.6  &  $33.0\pm 1.5$  &  $0.224\pm 0.009$  & ~51.85  &  15  &
$0.20^{+0.25}_{-0.20}$  \\
 188.6  &  $28.1\pm 1.2$  &  $0.245\pm 0.007$  & 154.00  &  52  &  
$0.33^{+0.18}_{-0.16}$  \\ 
\hline\hline
\end{tabular}
\end{center}
\caption{Performance of the $\evqq$ event selection at the
  two centre-of-mass energies considered in the analysis.}
\label{tab:sigqq}
\end {table}

%A candidate event for $\evqq$ is shown in Figure~\ref{fig:evhad}.

\subsection{Selection of leptonic events}
\label{sec:Wev_lept}

The experimental signature of the leptonic channel 
$e^+e^-\to e^- \bar{\nu}_e l^+ \nu_l$ is the presence of a 
high energy lepton accompanied by a large missing 
momentum and no other significant energy deposition in the detector. 
The analysis was optimised for final state leptons that
are electrons or muons.
%LASTCOR (begin)
%LASTCOR , and its sensitivity to $\evtv$ is only marginal. 
%LASTCOR In the $e^+e^-\to e^- \bar{\nu}_e l^+ \nu_l$ $(l \neq e)$ channel, the
%LASTCOR measured cross-section includes the contribution of both the muon and
%LASTCOR the tau final states; in the electron channel, 
In both channels,
%LASTCOR (end)
the contribution from $\evtv$ events was considered as part of the background.

The main backgrounds for the leptonic channel are the radiative
production of two leptons $\eell(\gamma)$, $\eeww$ events and two-photon
collisions. 

Events were selected if exactly one well measured charged particle was 
reconstructed. The quality of the track measurement was assessed as
follows: 
\begin{itemize}
\item relative error on the momentum, $\Delta{p}/p$, smaller than $100\%$;
\item track length greater than 20~cm;
\item polar angle $\theta$ between $10^\circ$ and $170^\circ$;
\item impact parameter in the transverse plane, $|IP_{R\phi}|$, smaller
  than 4~cm, and that along the beam direction, $|IP_z|$, smaller than 3~cm~$/\sin\theta$.
\end{itemize}
Loose identification criteria were applied, requiring associated hits in the
muon chambers or a significant energy deposition in the
electromagnetic calorimeter.
For electrons, the acceptance was
restricted to the barrel region, $|\cos\theta|<0.72$, and the best
determination of the electron energy was estimated by combining the
momentum measurement from the tracking devices and the calorimetric
energy. 
Any other energy deposit in the detector not related to the lepton candidate
was required not to exceed 2~GeV. In addition, the
presence of tracks not fulfilling the quality criteria
listed above was used to veto the event.
The acceptance was restricted to the kinematic region of $W$ decays by
requiring the lepton momentum to lie below 45\% of
$\sqrt{s}$ and its transverse momentum to exceed 12\% of $\sqrt{s}$.

A large residual contamination was still present, due to cosmic ray events
in the muon channel and to Compton scattering in the electron
channel. The former were suppressed by tightening the selections on the
track impact parameters to $|IP_{R\phi}|<0.2$~cm and
$|IP_z|<2$~cm for the muons. 
Compton events can mimic the $W^+\to{e^+}\nu_e$ signal when
the photon balancing the electron in the transverse plane is lost in
the dead region between the barrel and forward electromagnetic
calorimeters. Therefore events were rejected if a signal was found in
the hermeticity counters at an azimuthal angle larger than 90$^\circ$ 
from the electron.

\begin{figure}[htb]
\begin{center}
\epsfig{file=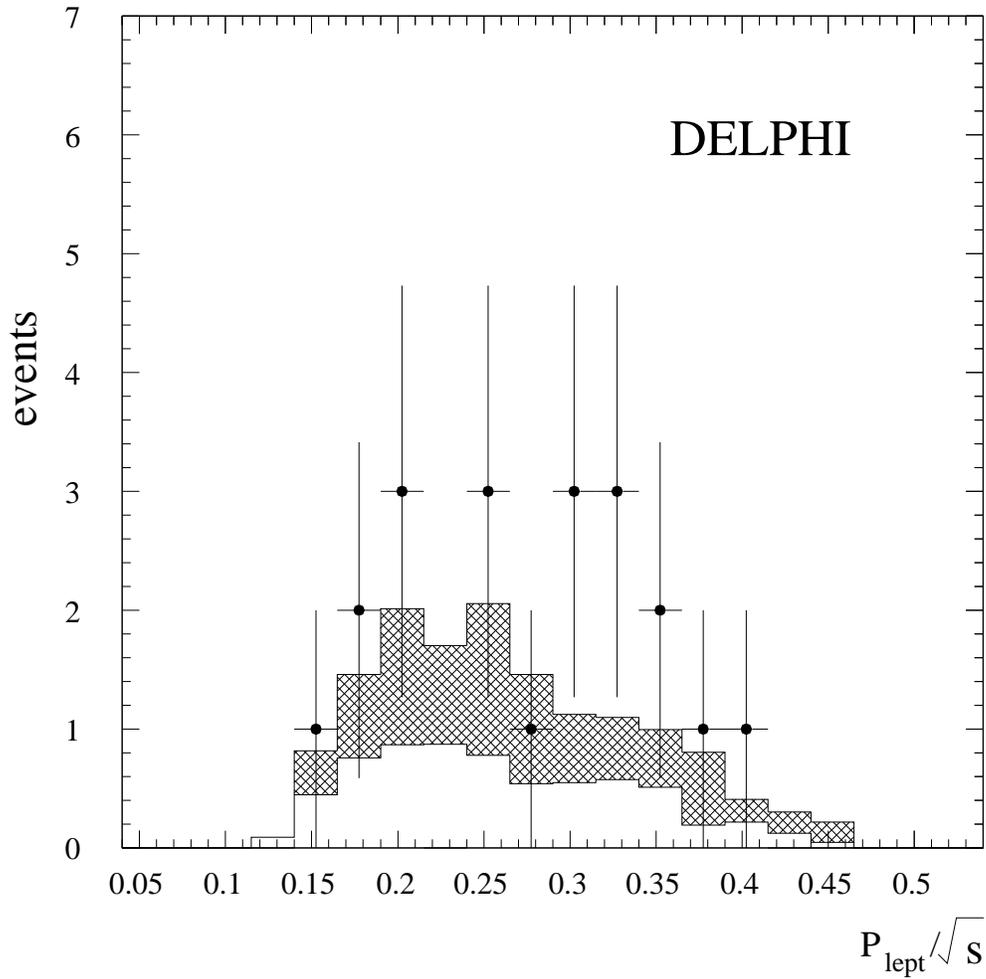,width=13cm}
\end{center}
\caption{$\wev$ channel $(W \to  l^+ \nu_l)$ at  $\sqrt{s}=189$~GeV.
  Momentum distributions of the lepton $l^+$ in real data
  (points with error bars) and in the simulation
  (histograms) for the events selected at the end of the analysis. The
  open area represents the single-$W$ signal, the cross-hatched histogram is
  the background expectation. }
\label{fig:plep}
\end{figure}

Figure~\ref{fig:plep} shows the momentum distribution of single
leptons in data and simulation at 189 GeV.
The performance of the analysis at the two centre-of-mass energy values and
the results obtained are reported in Table~\ref{tab:evlv}.

%LASTCORDB (begin)
\begin{table}[tb]
\begin{center}
\vspace{0.5cm}
\begin{tabular}{|lc|c|c|c|c||c|}
\hline\hline
\multicolumn{2}{|c|}{$\sqrt{s}$ (GeV)} & Eff. on $\mu$ (\%) &   
$\sigma_{bkg}$ (pb) & $\lu$ (\pbi) & $N_{data}$ & $\sigma_{e\nu \mu \nu}$ (pb)  \\
\hline \hline
 $l=\mu$ & 182.6   &  $70.8\pm 1.0$  &    $0.013\pm 0.002$ &
 ~51.85  &  ~6  & $0.147^{+0.076}_{-0.058}$ \\
 ~          & 188.6   &  $63.1\pm 1.0$  &    $0.012\pm 0.002$  &   
 153.45  &  ~8 & $0.062^{+0.033}_{-0.026}$ \\  
\hline \hline
\multicolumn{2}{|c|}{$\sqrt{s}$ (GeV)}  &  Eff. on $e$ (\%)   & $\sigma_{bkg}$ (pb) &
     $\lu$ (\pbi) &   $N_{data}$        &   $\sigma_{e\nu e\nu}$ (pb)  \\
\hline \hline
 $l=e$ & 182.6     &  $83.4\pm 3.2$  &     $0.038\pm 0.008$ &
 ~51.85  &  ~3  & $ 0.024^{+0.048}_{-0.024}$ \\ 
 ~   & 188.6     &  $81.1\pm 0.9$  &     $0.043\pm 0.008$  &
 153.45  &  12  & $ 0.044^{+0.031}_{-0.026}$ \\
\hline\hline
\end{tabular}
\end{center}
\caption{Performance of the $e^-\bar{\nu}_e \mu^+ \nu_{\mu}$  and $e\nu e\nu$
  event selection at the two centre-of-mass energies considered in the analysis. }
\label{tab:evlv}
\end {table}
%LASTCORDB (end)

%LASTCOR \begin{table}[tb]
%LASTCOR \begin{center}
%LASTCOR \vspace{0.5cm}
%LASTCOR \begin{tabular}{|lc|c|c|c|c|c||c|}
%LASTCOR \hline\hline
%LASTCOR \multicolumn{2}{|c|}{$\sqrt{s}$ (GeV)} & Eff. on $\mu$ (\%) &  Eff. on $\tau$ (\%) & 
%LASTCOR $\sigma_{bkg}$ (pb) & $\lu$ (\pbi) & $N_{data}$ & $\sigma_{e\nu l\nu}$ (pb)  \\
%LASTCOR \hline \hline
%LASTCOR  $l \neq e$ & 182.6   &  $70.8\pm 1.0$  &  $2.5\pm 0.5$  &  $0.012\pm 0.002$ &
%LASTCOR  ~51.85  &  ~6  & $0.29^{+0.15}_{-0.11}$ \\
%LASTCOR  ~          & 188.6   &  $63.1\pm 1.0$  &  $2.5\pm 0.5$  &  $0.011\pm 0.002$  &   
%LASTCOR  153.45  &  ~8 & $0.13^{+0.06}_{-0.05}$ \\  
%LASTCOR \hline \hline
%LASTCOR \multicolumn{2}{|c|}{$\sqrt{s}$ (GeV)}  &  Eff. on $e$ (\%)  &   -  & $\sigma_{bkg}$ (pb) &
%LASTCOR      $\lu$ (\pbi) &   $N_{data}$        &   $\sigma_{e\nu e\nu}$ (pb)  \\
%LASTCOR \hline \hline
%LASTCOR  $l=e$ & 182.6     &  $83.4\pm 3.2$  &   -   &  $0.038\pm 0.008$ &
%LASTCOR  ~51.85  &  ~3  & $ 0.024^{+0.048}_{-0.024}$ \\ 
%LASTCOR  ~   & 188.6     &  $81.1\pm 0.9$  &   -   &  $0.043\pm 0.008$  &
%LASTCOR  153.45  &  12  & $ 0.044^{+0.031}_{-0.026}$ \\
%LASTCOR \hline\hline
%LASTCOR \end{tabular}
%LASTCOR \end{center}
%LASTCOR \caption{Performance of the $e^-\bar{\nu}_e l^+ \nu_l$ $(l\neq e)$ and $e\nu e\nu$
%LASTCOR   event selection at the two centre-of-mass energies considered in the analysis. }
%LASTCOR \label{tab:evlv}
%LASTCOR \end {table}

\subsection{Study of systematic uncertainties}

The main source of systematic error is the limited simulation
statistics, both for the signal and for the background.
However this has little influence on the accuracy of the measurement,
since the error is dominated by the real data statistics.

Possible inaccuracies in the modelling of background processes were
evaluated by comparing different Monte Carlo generators. The only
notable effect was found in the $\qq(\gamma)$ channel, where the
background estimate to the hadronic selection evaluated with the
ARIADNE~\cite{ariadne} event generator was found to be 
$6.0\pm0.6$ events at 189~GeV. 
The difference from the value obtained from the \mbox{PYTHIA} samples ($5.1\pm 0.3$, 
see Table~\ref{tab:nevts}) was considered as a systematic uncertainty.

%The value of the $\eeww$ cross-section measured by the LEP experiments
%on the 189~GeV data~\cite{sigww} was lower than the Standard Model
%expectation. This discrepancy was taken into account as a systematic
%uncertainty on the contamination due to $WW$ events.

The total systematic error on the background cross-section, mainly due to the
effects listed above, amounts approximately to $\pm5\%$ in the $q\bar{q}'$ channel 
and to $\pm 20\%$ in each of the leptonic channels 
(see Tables~\ref{tab:sigqq} and~\ref{tab:evlv}, for the part due only to the Monte Carlo statistics).

From a comparison of dimuon events in data and simulation, the
tracking efficiency, $\varepsilon_{track}$,
of {\mbox DELPHI} was found to be overestimated by
0.5\% in the simulation. This difference was assumed as systematic error.
This has a negligible effect on the
background, while it affects the selection efficiency of the signal
for leptonic decays of the $W$.

The luminosity is known with a total error of $\pm0.6\%$. 

The effect of the uncertainties listed above on the measurement of the
$e\nu f\bar{f}'$ and $e\nu{e\nu}$
cross-sections at $\sqrt{s}$=189~GeV are given in 
Table~\ref{tab:syst}. The total systematic error, obtained from the
sum in quadrature of the individual contributions, is at the level of $\pm10\%$
for $e^-\bar{\nu}_e q\bar{q}'$ and about $\pm25\%$ in the case of $e\nu{e\nu}$.
For the measurement at 183~GeV, the same relative error was assumed.

%LASTCOR (begin)
\begin{table}[tb]
\begin{center}
\vspace{0.5cm}
\begin{tabular}{|l|c|c|}
\hline\hline
 Systematic effect  &   Error on $\sigma_{e\nu f\bar{f}'}$ (pb)  & 
 Error on $\sigma_{\evev}$ (pb) \\  \hline\hline
 $\Delta\sigma_{bkg}$ ($\evqq$) $\pm  5\%$             &   0.041  & -     \\
 $\Delta\sigma_{bkg}$ ($\evev$) $\pm 20\%$          &    -     & 0.0106  \\
 $\Delta\sigma_{bkg}$ ($\evmv$) $\pm 20\%$      &   0.004  &  -    \\ 
\hline
 $\Delta\varepsilon$ ($\evqq$) due to simul. stat.     &   0.014  & -     \\
 $\Delta\varepsilon$ ($\evlv$) due to simul. stat     &   0.001  & 0.0005  \\ 
\hline
 $\Delta\varepsilon$ ($\evlv$) due to $\varepsilon_{track}$    
                                                     &   0.001  & 0.0002  \\ 
\hline
 Luminosity  $\pm 0.6\%$                             &   0.008  & 0.0006  \\
 \hline\hline
 Total                                               &   0.044  & 0.0106 \\
 \hline\hline
\end{tabular}
\end{center}
\caption{Contributions to the systematic uncertainty 
on the $e\nu f\bar{f}'$ and $e\nu e{\nu}$ cross-sections at $\sqrt{s}=189$~GeV.}
\label{tab:syst}
\end {table}
%LASTCOR (end)

%LASTCOR \begin{table}[tb]
%LASTCOR \begin{center}
%LASTCOR \vspace{0.5cm}
%LASTCOR \begin{tabular}{|l|c|c|}
%LASTCOR \hline\hline
%LASTCOR  Systematic effect  &   Error on $\sigma_{e\nu f\bar{f}'}$ (pb)  & 
%LASTCOR  Error on $\sigma_{\evev}$ (pb) \\  \hline\hline
%LASTCOR  $\Delta\sigma_{bkg}$ ($\evqq$) $\pm  5\%$             &   0.041  & -     \\
%LASTCOR  $\Delta\sigma_{bkg}$ ($\evev$) $\pm 20\%$          &    -     & 0.0106  \\
%LASTCOR  $\Delta\sigma_{bkg}$ ($\evmv$) $\pm 20\%$      &   0.006  &  -    \\ 
%LASTCOR \hline
%LASTCOR  $\Delta\varepsilon$ ($\evqq$) due to simul. stat.     &   0.014  & -     \\
%LASTCOR  $\Delta\varepsilon$ ($\evlv$) due to simul. stat     &   0.004  & 0.0005  \\ 
%LASTCOR \hline
%LASTCOR  $\Delta\varepsilon$ ($\evlv$) due to $\varepsilon_{track}$    
%LASTCOR                                                      &   0.001  & 0.0002  \\ 
%LASTCOR \hline
%LASTCOR  Luminosity  $\pm 0.6\%$                             &   0.008  & 0.0006  \\
%LASTCOR  \hline\hline
%LASTCOR  Total                                               &   0.045  & 0.0106 \\
%LASTCOR  \hline\hline
%LASTCOR \end{tabular}
%LASTCOR \end{center}
%LASTCOR \caption{Contributions to the systematic uncertainty 
%LASTCOR on the $e\nu f\bar{f}'$ and $e\nu e{\nu}$ cross-sections at $\sqrt{s}=189$~GeV.}
%LASTCOR \label{tab:syst}
%LASTCOR \end {table}

\section{Single-\mbox{\boldmath $Z$} analysis}
In the single $\gs/Z$ analysis, decays of the vector boson into hadronic and
$\mm$ final states were considered. Both final states are
characterized by an electron scattered at large angle
with respect to the incoming direction.  
The other electron, lost in the beam pipe, results in a
missing momentum pointing along the beam line direction. 
Instead of attempting to separate the  $\gs ee$ from the $Zee$
contributions it was preferred, like in~\cite{opalZee}, 
to determine the cross-sections in two separate ranges of the invariant
mass, $m_{f\bar{f}}$, of the final system: from 15 to 60 GeV/$c^2$ and above
60 GeV/$c^2$. The value of 60 GeV/$c^2$ was chosen as it represents
about the minimum of the differential $m_{f\bar{f}}$ distribution.

\subsection{Selection of hadronic events}

The experimental signature of these events consists of a pair of jets
produced in the hadronic decay of the $\gs/Z$ recoiling against an
electron. 
To maximize the sensitivity of the analysis in the widest possible range of 
invariant masses of the $\gs/Z$, the event selection was performed in three steps:
\begin{enumerate}
\item a loose preselection of hadronic events;
\item the identification of an isolated electron;
\item the final selection of signal events, optimized differently in
  two ranges of the invariant mass of the hadronic system, $m_{q\qbar}$,
  according to the most relevant background process in each region.
\end{enumerate}

\begin{table}[tb]
\begin{center}
\vspace{0.5cm}
\begin{tabular}{|l|c|c|c|c|c|c|r|}
\hline\hline
       & $\gs/Zee$     & $WW$           &  $Z(\gamma)$ & $\gamma\gamma$ &  Others        &     Total MC  & Data
\\ \hline\hline                                                               
Preselection                                                                   
      & 179.9           & 1046.7        & 3156.5       & 887.6          &  103.3         &       5374.0  & 5812      
\\ \hline                                                                     
$e$ ident.                                                            
      & ~95.7           & ~118.3        & ~~64.0       & 126.5          &  ~~4.5         &        ~409.0 & 400
\\ \hline                                                                     
Signal selection
       & 37.3 $\pm$ 2.5 & 3.2 $\pm$ 0.4 & 7.0 $\pm$ 0.6& 6.9 $\pm$ 2.1  &  0.4 $\pm$ 0.1 & 54.8 $\pm$ 3.3 & 51
\\ \hline\hline
\end{tabular}
\end{center}
\caption{Number of events expected from the contribution of different
       channels and observed in the data at different stages of the
       $\gs/Zee$ selection at $\sqrt{s} = 189$~GeV in the hadronic channel. 
       The number of
       expected $\gs/Zee$ events has been computed using  a simulation
       sample generated with \mbox{GRC4F}.
       The column labelled ``Others'' includes Bhabha events and other
       four-fermion 
       processes, namely $\wev$ and $\gs/Zee$ with fully leptonic
       final state. 
       Details on the selection are provided in Section~4.1.}
\label{tab:Zee1}
\end {table}

%\paragraph{{\bf Preselection:}} 
\noindent The preselection of hadronic events consisted of the following
requirements:
\begin{itemize}
\item at least five charged particles in the event with at least one 
  in the Time Projection Chamber, the main {\mbox DELPHI} tracking detector, 
  with a measured transverse momentum larger than 2.5~GeV/$c$;
\item in events with more than one electromagnetic shower, 
  the energy of the second most energetic one was required to be less than
  $0.6 E_{beam}$ in order to reject Bhabha events.
\end{itemize}

%\paragraph{{\bf Electron identification:}}
\noindent The electron candidates were selected by requiring energy  
depositions in the calorimeter $E_e > 4$~GeV, with an associated charged
particle and in the angular acceptance  $|\cos \theta_e| < 0.985$. 
Moreover they had to satisfy the following isolation criteria:
\begin{itemize}
\item their angle, $\alpha$, with respect to the closest particle
  with momentum $p > 0.5$~GeV/$c$ had to lie in the range $15\deg <
  \alpha < 170\deg$;
\item their angle  with respect to the second closest particle, 
  with $p > 0.5$~GeV/$c$, had to be greater than $40\deg$. 
\end{itemize}
Electrons from conversions or from decays were further reduced
by requiring their impact parameters with respect to the primary
interaction vertex to be $|IP_{R\phi}|<0.35$~cm in the transverse
plane and  $|IP_{z}|<1$~cm along the beam line.

The charged and neutral particles were then clustered into two jets with the Durham
algorithm, excluding the tag electron and rejecting events for which 
$D^{join}_{3 \to 2} < 10^{-4}$. A kinematic fit of the event was then 
performed assuming a topology of signal events with two jets, a
visible electron and one lost along the beam line. 
The four-momentum of the invisible electron was chosen to be 
$(0,0,Q_{e}E,E)$ with $Q_{e}$ the charge of the tagged
electron\footnote{The {\mbox DELPHI} reference frame has the $z$ axis
  oriented along the incoming $e^-$ beam.}. 
Fits with a $\chi^2$ probability smaller than $10^{-5}$ were
rejected.

\begin{figure}[htb]
\begin{center}
\epsfig{file=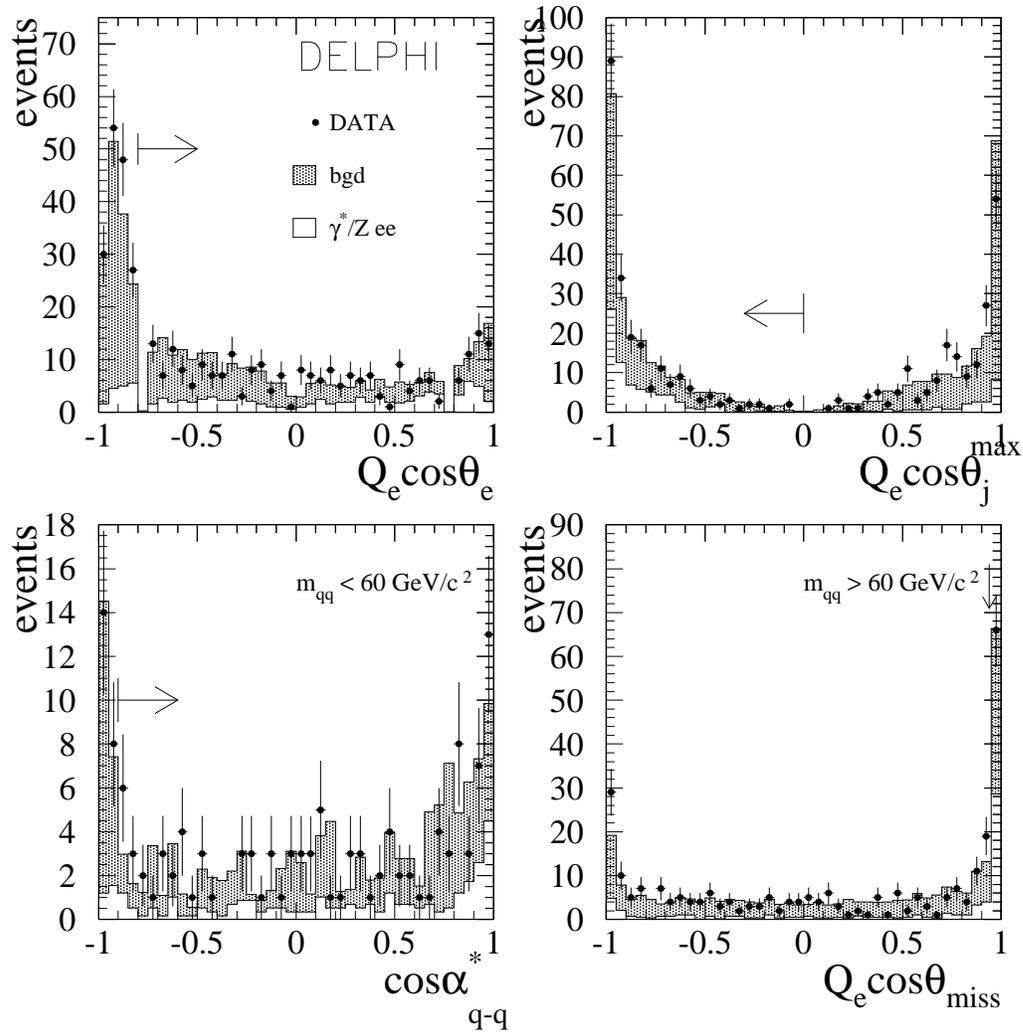,width=15cm}
\end{center}
\caption{$\gs/Zee$ channel $(\gs/Z \to q\bar{q})$ at $\sqrt{s} = 189$ GeV.
  Distributions of the variables used for the signal definition
  after the ``electron identification'' step (see Section~4.1), in real data
  (points with error bars) and in the simulation (histograms). 
  The arrows indicate the value of the cut on each variable. 
  The top plots show discriminant variables used for the signal selection
  in the overall $m_{q\qbar}$ spectrum. 
  The bottom ones show the variables used for the different selections
  in the low (left) and high (right) invariant mass region of the
  hadronic system.} 
\label{fig:ZeeCuts}
\end{figure}

%\paragraph{{\bf Signal selection:}}
 The final selection of signal events was then performed  
using the variables after the constrained fit.
It was required that:
\begin{itemize}
\item $Q_{e} \cos \theta_{e} > -0.8$ with $\theta_e$ being the
  polar angle of the tagged electron;
\item $Q_{e} \cos \theta_{j}^{max} < 0$ with $\theta_{j}^{max}$ being
  the polar angle of the jet closest to the beam line.
\end{itemize}
Two different sets of cuts were then applied in distinct regions of $m_{q\qbar}$. \\
For $m_{q\qbar}<60$~GeV/$c^2$, where the dominant background consisted of
two-photon events:
\begin{itemize}
\item $\cos \alpha_{q\qbar}^{*} > -0.9 $ with
  $\alpha_{q\qbar}^{*}$ being the angle between the two jets in the  
  electron-$\gs/Z$ rest frame;
\item $Q_{e} \cos \theta_{e} < 0.9$ or $E_{e}<0.75 E_{beam}$.
\end{itemize}
For $m_{q\qbar}>60$~GeV/$c^2$, where the dominant background consisted of
 $WW$ events:
\begin{itemize}
\item $Q_{e} \cos \theta_{miss} > 0.95$ with $\theta_{miss}$ being
  the polar angle of the missing momentum computed
  before the kinematic fit;
\item $Q_{e} \cos \theta_{j}^{max} > -0.985$.
\end{itemize}
The distributions of these variables after the electron identification
cuts are shown in Figure~\ref{fig:ZeeCuts} 
for the real and simulated data. 
The numbers of selected events in the data and the expected 
contributions from the different backgrounds 
after each selection step are shown
in Table~\ref{tab:Zee1}. 

\begin{figure}[htb]
\begin{center}
\epsfig{file=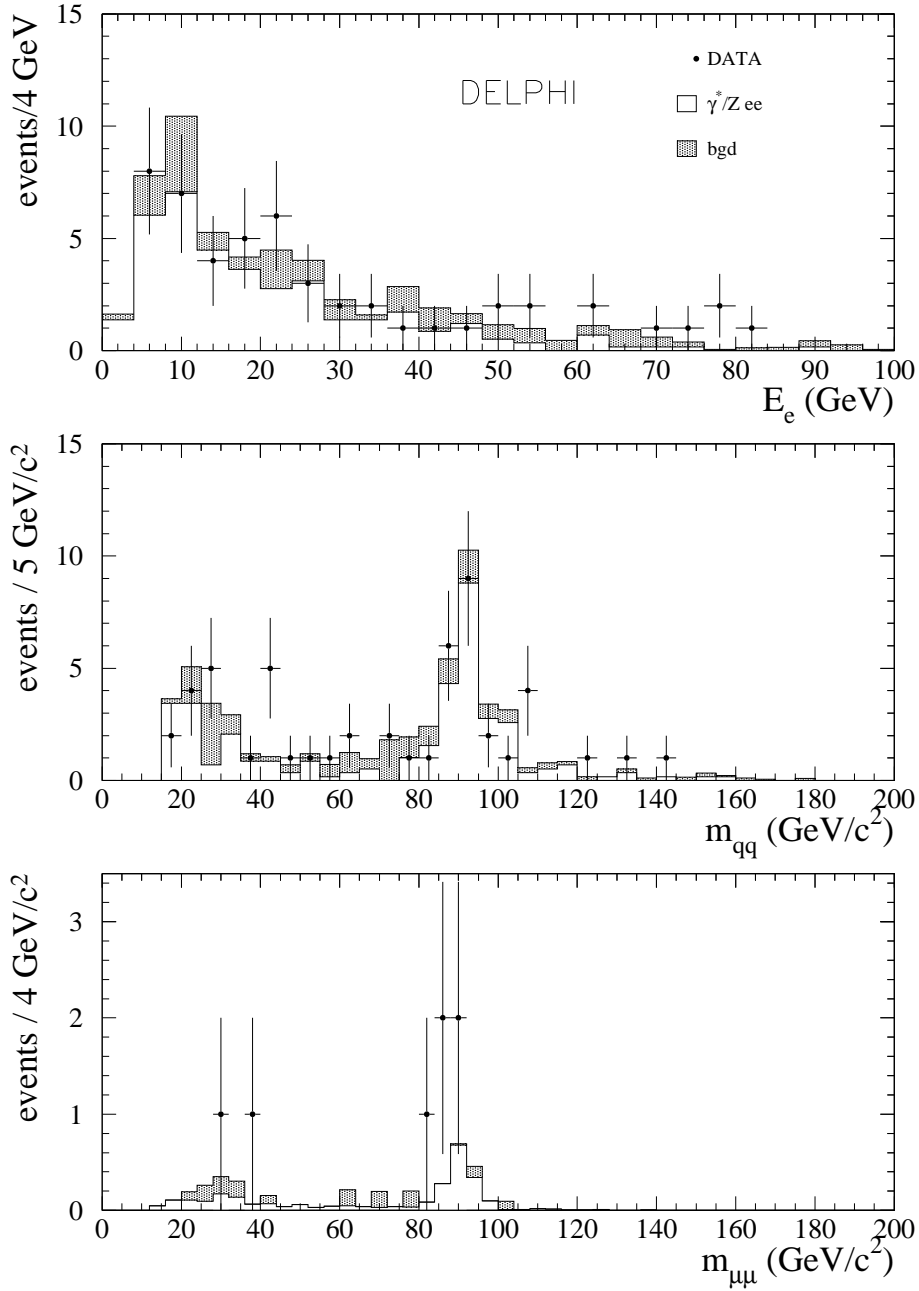,width=13cm}
\end{center}
\caption{$\gs/Zee$ channel at $\sqrt{s} = 189$ GeV. Energy spectrum of
  the tag electron {\it (top)} 
  and invariant mass distribution of $\gs/Z$ system 
  {\it (centre)} in real data (points with error bars) and in the
  simulation (histograms) in the case of hadronic final states.
  Invariant mass distribution of the $\gs/Z$ system 
  {\it (bottom)} in the case of $\mm$ final states.}
\label{fig:ZeeMff}
\end{figure}

The efficiency of the selection on the signal, the expected background
and the number of selected events in the data at the two centre-of-mass
energies are reported in Table~\ref{tab:ZeePerf}, together with the evaluated 
cross-section. The distribution of the invariant mass of the hadronic
system and the energy spectrum of the tag electron after the
kinematic fit are shown in Figure~\ref{fig:ZeeMff}.
The peak in the invariant mass distribution around the $Z$ mass 
corresponds to events for which the contribution of the $Zee$ process
is dominant.

\begin{table}[tb]
\begin{center}
\vspace{0.5cm}
\begin{tabular}{|c|c|c|c|c|c|c|}
\hline\hline
$\sqrt{s}$ & $\gs/ Z \to \qq$       &  Eff.  &   $\sigma_{bgd}$ & $\lu$ &   $N_{data}$ &   $\sigma$ \\
 (GeV)     & mass range (GeV/$c^2$) & (\%)  &  (pb)         & (pb$^{-1}$) & & (pb) \\
\hline \hline
 182.6  & $15 < m_{q\qbar} < 60$ &  19.2 $\pm$  2.0 &  0.05
 $\pm$ 0.01 & ~51.9 & ~6 & $0.33^{+0.28}_{-0.21}$  \\
        & $~~~m_{q\qbar} > 60$            &  14.2 $\pm$  1.2 &  0.05
 $\pm$ 0.01 & ~51.9 & 15 & $1.66^{+0.57}_{-0.48}$ \\  
\hline 
188.6  & $15 < m_{q\qbar} < 60$ &  20.1 $\pm$  2.1 &  0.05
 $\pm$ 0.01 & 154.3 & 20 & $0.41^{+0.16}_{-0.13}$ \\  
       & $~~~m_{q\qbar} > 60$  &  16.9 $\pm$  1.3 &  0.07
 $\pm$ 0.01 & 154.3 & 31 & $0.79^{+0.23}_{-0.20}$ \\
\hline\hline
$\sqrt{s}$ & $\gs/ Z \to \mm$       &  Eff.  &   $\sigma_{bgd}$ & $\lu$ &   $N_{data}$ &   $\sigma$ \\
 (GeV)     & mass range (GeV/$c^2$) & (\%)  &  (pb)         & (pb$^{-1}$) & & (pb) \\
\hline \hline
 182.6  & $15 < m_{\mm} < 60$ &  ~5.4 $\pm$ 0.3  &  0.004
 $\pm$ 0.002 & ~51.9 &  0 & -  \\
        & $~~~m_{\mm} > 60$            &  33.8 $\pm$ 1.5 &  0.004
 $\pm$ 0.001 & ~51.9 & 1 & -  \\  
\hline
188.6 & $15 < m_{\mm} < 60$ &  ~5.4 $\pm$ 0.3 &  0.004
 $\pm$ 0.002 & 154.3 & 2 & $0.154^{+0.206}_{-0.129}$ \\  
      & $~~~m_{\mm} > 60$  &  33.8 $\pm$ 1.5 &  0.005
 $\pm$ 0.002 & 154.3 & 5 & $0.080^{+0.048}_{-0.036}$ \\
\hline\hline
\end{tabular}
\end{center}
\caption{Performance of the $\gs/Zee$ event selection at the
  two centre-of-mass energies considered in the analysis.}
\label{tab:ZeePerf}
\end {table}

\subsection{Selection of leptonic events}

The search was restricted to events with $\gs/Z$ going 
into a $\mu^+\mu^-$ pair. The general features are exactly the same as
for the hadronic channel with jets replaced by muons.
Thus a three-track signature, of two high momentum muons and one
$e^+$ or $e^-$, scattered at large angle, is expected in the detector.
After a common preselection and lepton identification, 
the analysis was tuned separately for two kinematic regions:
$m_{\mm} > 60$~GeV/$c^2$ and $15 <m_{\mm}< 60$~GeV/$c^2$. 
The signal selection criteria on angular distributions were similar 
to those used in the hadronic channel.

%\paragraph{{\bf Preselection:}}
In the preselection the event was required to have exactly three tracks
fulfilling the following criteria:
\begin{itemize}
\item fractional error on the momentum $\Delta p/p < 50 \%$,
\item impact parameter in the transverse plane  $|IP_{R\phi}| < 0.5$~cm 
         and along the beam direction $|IP_{z}| < 3$~cm;
\item at least one associated hit in the Vertex Detector.
\end{itemize}
The sum of the charges of the three particles was required to
be $\pm 1$.
Possible photon conversions were removed according to the standard 
DELPHI procedure described in~\cite{DELPHIperf}.
The minimum opening angle of any track pair had to be larger than
$5^\circ$. 

%\paragraph{{\bf Lepton identification:}}
Since the event topology is clean, the particle identification 
required at least two tracks to be identified as leptons
($\mu$ or $e$) and at least one of them to be a muon. The momentum of
the electron had to be greater than 4~GeV/$c$. For muon identification
the loose 
criteria were applied as in the case of single-$W$ production 
(see Section~\ref{sec:Wev_lept}).
The flavour of the possible unidentified track was inferred from
partial information taking into account the combination of the charges of the 
observed particles.
In the case of $\mu^+x^-e^{\pm}$ or $x^+\mu^-e^{\pm}$, the unidentified track 
$x$ was treated as $\mu$. For $\mu^+\mu^-x^{\pm}$ the track $x$ was taken 
as $e^{\pm}$. In this way the loss of efficiency due to electron 
identification was minimal.

The data reduction factor of the preselection was large.
%For both centre-of-mass energies 44 events were preselected with 
%$44.8 \pm 1.7$ events expected. 
At $\sqrt{s} = 189$~GeV, 35 events were preselected with 
$33.6 \pm 1.6$ events expected. 
At this stage the majority of events came 
from the $\gamma\gamma \to \mu^+\mu^-$ process
(see Table~\ref{bgsplit}). The other non-zero
contributions came from the following sources (ordered by decreasing 
significance) : $e^+e^- \to \mu^+\mu^-(\gamma)$, $e^+e^- \to ZZ$, 
$e^+e^- \to l_1^+l_1^-l_2^+l_2^-~(l_1,l_2=e,\mu,\tau)$, 
$\gs/Ze^+e^- \to \tau^+\tau^-e^+e^-$, $e^+e^- \to W^+W^-$ and 
$e^+e^- \to \mu^+\mu^-q\bar{q}$.
\begin{table}[tb]
\begin{center}
\begin{tabular}
{|c|c|c|c|c|c|}
\hline
\hline
& $\gs/Zee$ & $\gamma\gamma \to \mu^+\mu^-$ & Others & Total MC & Data \\
\hline
\hline
Preselection     & $5.9\pm 0.1$ & $23.2\pm 1.4$ & $4.5\pm 0.7$ & $33.6\pm1.6$ & 35  \\ 
\hline
Final selection  & $2.73\pm 0.10$  & $1.25\pm 0.32$ & $0.14\pm 0.12$ & $4.12\pm 0.36$ & 7 \\
\hline
\hline
\end{tabular}
\end{center}
\caption[.]{Number of events expected from the contribution of different
       channels and observed in the data at different stages of the
       $\gs/Zee$ selection at $\sqrt{s} = 189$~GeV in the leptonic channel. 
The column ``Others'' shows the numbers for two or four fermion background 
processes excluding $\gamma\gamma \to \mu^+\mu^-$.}
\label{bgsplit}
\end{table}

%\paragraph{{\bf Signal selection:}}
The final selection of signal events was similar for the two $m_{\mm}$ ranges.
%The angular distributions was marked by the charge $Q_e$ of the observed 
%electron track.
The allowed angular ranges for the direction of the $Z/\gamma^*$
momentum and missing momentum were defined by the following
conditions, in which $Q_e$ represents the charge of the observed electron:
\begin{itemize}
\item
$Q_e \cos\theta_{\mu^+\mu^-}~>~-0.8$~~~with $\theta_{\mu^+\mu^-}$ 
being the polar angle of the $\mu^+\mu^-$ system
\item
$Q_e \cos\theta_{miss}~>~0.8$~~~~~with $\theta_{miss}$ being the polar angle 
of the missing momentum.
\end{itemize}
The final selection was dependent on the $\mm$ invariant mass:
\begin{itemize}
\item
$Q_e \cos\theta_{e}~>~-0.8$~~~for $m_{\mm}$ greater than 60~GeV/$c^2$
\item
$Q_e \cos\theta_{e}~>~-0.7$~~~for $m_{\mm}$ between 15 and 60~GeV/$c^2$.
\end{itemize}
The stronger condition for the low invariant mass region was to 
reduce the background from the
$\gamma\gamma \to \mu^+\mu^-$ process further. The sum of all other sources
such as two- or four-fermion production is an order of magnitude
smaller after the final selection.

At 183~GeV in the high invariant mass $m_{\mm}$ region, 
one event was found in the data with $0.8\pm 0.1$ event expected.
For the low mass region no event was observed in the data 
and $0.6\pm0.1$ event was expected. Due to the low statistics, 
the value of the cross-section was not derived.
At 189~GeV, where the integrated luminosity was three times greater, 
5 events ($2.5\pm0.3$ predicted) and 2 events ($1.6\pm0.2$ predicted)
were selected in the high and low mass regions, respectively.
The spectrum of the $m_{\mm}$ invariant mass after the kinematic fit
is shown in Figure~\ref{fig:ZeeMff} for the data at 189~GeV.
A kinematic fit, assuming the lost electron along the beam line and no 
missing momentum in the transverse plane, was applied to improve the mass 
resolution.

The efficiency of the selection on the signal, the expected background
and the number of selected events in the data at the two
centre-of-mass energies are reported in Table~\ref{tab:ZeePerf}
together with the evaluated cross-sections.

\subsection{Systematic uncertainties}

\begin{table}[tb]
\begin{center}
\vspace{0.5cm}
\begin{tabular}{|l|c|c|}
\hline\hline
 Systematic effect  &   \multicolumn{2}{|c|}{Error on $\sigma$ (pb)}  \\
\hline \hline
& $15 < m_{q\qbar} < 60$~GeV/$c^2$ & $m_{q\qbar} > 60$
~GeV/$c^2$ \\
\hline
 $\Delta\varepsilon_e$                                 &  0.034 &  0.064 \\ 
\hline
 $\Delta\sigma_{bkg}$ ($\gamma \gamma$) $\pm 20\%$     &  0.030  & 0.015 \\
\hline
 $\Delta\varepsilon$  due to simul. stat               &  0.043  & 0.060 \\ 
 $\Delta\sigma_{bkg}$ due to simul. stat.              &  0.058  & 0.051 \\
\hline
 Luminosity  $\pm 0.6\%$                               &  0.003  & 0.005 \\
 \hline\hline
 Total                                                 &  0.085  & 0.103 \\
 \hline\hline
\end{tabular}
\end{center}
\caption{Contributions to the systematic uncertainty 
 on the $\gs/Zee$ cross-sections in the hadronic channel, in the two ranges
 of invariant mass of the hadronic system, at $\sqrt{s}=189$~GeV.}
\label{tab:ZeeSyst}
\end {table}
\begin{table}[tb]
\begin{center}
\vspace{0.5cm}
\begin{tabular}{|l|c|c|}
\hline\hline
 Systematic effect  &   \multicolumn{2}{|c|}{Error on $\sigma$ (pb)}  \\
\hline \hline
& $15 < m_{\mm} < 60$~GeV/$c^2$ & $m_{\mm} > 60$
~GeV/$c^2$ \\
\hline
 $\Delta\sigma_{bkg}$ ($\gamma \gamma$) $\pm 20\%$     &  0.016  & 0.003 \\
\hline
 $\Delta\varepsilon$  due to simul. stat               &  0.010  & 0.004 \\ 
 $\Delta\sigma_{bkg}$ due to simul. stat.              &  0.028  & 0.005 \\
\hline
 Luminosity  $\pm 0.6\%$                               &  0.001  & 0.001 \\
 \hline\hline
 Total                                                 &  0.034  & 0.007 \\
 \hline\hline
\end{tabular}
\end{center}
\caption{Contributions to the systematic uncertainty 
 on the $\gs/Zee$ cross-sections in the leptonic channel, in the two ranges
 of invariant mass of the $\mu^+\mu^-$ system, at $\sqrt{s}=189$~GeV.}
\label{tab:ZeeSystmm}
\end {table}

In both channels the main systematic uncertainty came from the
limited simulation statistics available both for the signal and the background.
As in the case of the single-$W$ analysis the influence on the overall
error is limited since the measurement is dominated by the real data statistics.

Besides this, in the hadronic channel two other sources of systematic errors
were considered: the efficiency in the electron identification
procedure and the limited knowledge of the contribution from 
two-photon events which represents the largest background
component. 

The uncertainty on the efficiency of the electron identification was
estimated by comparing the number of selected events in the data and in
the simulation for a sample enriched in $WW$ events with at least one of
the two $W$'s decaying, directly or in cascade, into a final state
containing an electron. The same criteria for electron identification
and isolation were adopted as in the $Zee$ analysis. 
The relative difference in the efficiency was found to be $ \Delta
\varepsilon_{e}/\varepsilon_{e} = (6.7 \pm 8.2)\%$ 
where the error accounts both for the statistics and the
systematics due to the presence of about  11\% of background events
in the selected sample.
Conservatively, the error on the difference was used for
the computation of the systematic error.

A $\pm 20\%$ uncertainty on the $\gamma \gamma$ background was assumed, 
as determined from a study on single tag events for both investigated
final states.

The contributions of the different sources of systematics in the
hadronic channel at 189 GeV are summarized in Table~\ref{tab:ZeeSyst}.
The total systematic uncertainty amounts to $\pm21\%$ in the region 
$15 < m_{q\qbar}<60$~GeV/$c^2$ and to $\pm13\%$ for  
$m_{q\qbar}>60$~GeV/$c^2$.

The contributions of the different sources of systematics in the
leptonic channel at 189 GeV are summarized in Table~\ref{tab:ZeeSystmm}.
The total systematic uncertainties amount to $\pm22\%$ in the region 
$15 < m_{\mm}<60$~GeV/$c^2$ and to $\pm9\%$ for  
$m_{\mm}>60$~GeV/$c^2$.
%LASTCORDB (begin)
For the measurement at 183~GeV, the same relative error was assumed.
%LASTCORDB (end)

\section{Conclusions}

The cross-sections for single-$W$ production in the channels $\evqq$ and $\evlv$ $(l \neq e)$
%LASTCOR (begin)
, assuming $\mu-\tau$ universality, 
%LASTCOR (end)
have been measured in $e^+e^-$ collisions at 182.6 and 188.6~GeV centre-of-mass energies
by the {\mbox DELPHI} collaboration. 
%LASTCORDB (begin)
These cross-sections have been determined within a restricted phase-space (see Section~2).
%LASTCORDB (end)
The overall values are:
%LASTCOR (begin)
$$ \sigma_{e\nu f\bar{f}'} 
= 0.49~^{+0.27}_{-0.22} ~\mbox{(stat.)} \pm 0.04 ~\mbox{(syst.) pb} 
~~~~~~~~~(\sqrt{s} = 182.6~\mbox{GeV}), $$
$$ \sigma_{e\nu f\bar{f}'}
 = 0.45~^{+0.19}_{-0.16} ~\mbox{(stat.)} \pm 0.04 ~\mbox{(syst.) pb} 
~~~~~~~~~(\sqrt{s} = 188.6~\mbox{GeV}),  $$
in agreement with the Standard Model expectations of 0.37~pb and 0.41~pb, respectively.
%LASTCOR (end)

%LASTCOR $$ \sigma_{e\nu f\bar{f}'} 
%LASTCOR = 0.48~^{+0.29}_{-0.23} ~\mbox{(stat.)} \pm 0.05 ~\mbox{(syst.) pb} 
%LASTCOR ~~~~~~~~~(\sqrt{s} = 182.6~\mbox{GeV}), $$
%LASTCOR $$ \sigma_{e\nu f\bar{f}'}
%LASTCOR  = 0.46~^{+0.19}_{-0.17} ~\mbox{(stat.)} \pm 0.05 ~\mbox{(syst.) pb} 
%LASTCOR ~~~~~~~~~(\sqrt{s} = 188.6~\mbox{GeV}), $$
%LASTCOR in agreement with the Standard Model expectations of 0.37~pb and 0.41~pb, respectively.

In addition, the cross-sections for $\epem \to e\nu
e\nu$, which include contributions both from single-$W$
and from single-$Z$ with a large interference between the two 
processes, have been measured to be:
$$ \sigma (\epem\to e \nu e {\nu})
= 0.024~^{+0.048}_{-0.024} ~\mbox{(stat.)} \pm  0.006 ~\mbox{(syst.) pb} 
~~~~~~~~~(\sqrt{s} = 182.6~\mbox{GeV}), $$
$$ \sigma (\epem\to e \nu e {\nu})
 = 0.044~^{+0.031}_{-0.026} ~\mbox{(stat.)} \pm 0.011 ~\mbox{(syst.) pb} 
~~~~~~~~~(\sqrt{s} = 188.6~\mbox{GeV}), $$
in agreement with the Standard Model expectations of 0.041~pb and 0.046~pb, respectively.
In both cases, the theoretical predictions have been 
computed with the \mbox{GRC4F}~\cite{GRC4F} and 
\mbox{WPHACT}~\cite{WPHACT} programs for the signal phase space  
defined in equations (\ref{eq:cut1}) and (\ref{eq:cut2}). 

In the hadronic channel the cross-sections for single-$Z$ production at
$\sqrt{s} = 182.6$~GeV are:

\begin{displaymath}
\begin{array}{cr}
 \sigma
 = 0.33~^{+0.28}_{-0.21}~{\mathrm (stat.)} \pm 0.08~{\mathrm (syst.)~pb} &  15 < m_{q\qbar} < 60~{\mathrm GeV}/c^2, \\
\sigma 
 = 1.66~^{+0.57}_{-0.48}~{\mathrm (stat.)} \pm 0.21~{\mathrm (syst.)~pb} &  m_{q\qbar} > 60~{\mathrm GeV}/c^2,
\end{array}
\end{displaymath}

\noindent and at $\sqrt{s} = 188.6$~GeV:

\begin{displaymath}
\begin{array}{cr}
  \sigma 
 = 0.41~^{+0.16}_{-0.13}~{\mathrm (stat.)} \pm 0.09~{\mathrm (syst.)~pb} & 15 < m_{q\qbar} < 60~{\mathrm GeV}/c^2, \\
 \sigma 
 = 0.79~^{+0.23}_{-0.20}~{\mathrm (stat.)} \pm 0.10~{\mathrm (syst.)~pb} & m_{q\qbar} > 60~{\mathrm GeV}/c^2.
\end{array}
\end{displaymath}
\noindent
These values are found to be in agreement with Standard Model
predictions, computed with \mbox{GRC4F}, which at $\sqrt{s} =
182.6$~GeV are:

\begin{displaymath}
\begin{array}{cr}
 \sigma
 = 0.45~{\mathrm pb} &  15 < m_{q\qbar} < 60~{\mathrm GeV}/c^2, \\
\sigma 
 = 0.91~{\mathrm pb} &  m_{q\qbar} > 60~{\mathrm GeV}/c^2,
\end{array}
\end{displaymath}

\noindent and at $\sqrt{s} = 188.6$~GeV:

\begin{displaymath}
\begin{array}{cr}
 \sigma
 = 0.42~{\mathrm pb} &  15 < m_{q\qbar} < 60~{\mathrm  GeV}/c^2, \\
\sigma 
 = 0.94~{\mathrm pb} &  m_{q\qbar} > 60~{\mathrm GeV}/c^2.
\end{array}
\end{displaymath}
In the leptonic single-$Z$ channel the cross-sections were determined at $\sqrt{s} = 188.6$~GeV only:

\begin{displaymath}
\begin{array}{cr}
 \sigma
 = 0.15~^{+0.21}_{-0.13}~{\mathrm (stat.)} \pm 0.03~{\mathrm (syst.)~pb} &  15 < m_{\mm} < 60~{\mathrm GeV}/c^2, \\
\sigma 
 = 0.08~^{+0.05}_{-0.04}~{\mathrm (stat.)} \pm 0.01~{\mathrm (syst.)~pb} &  m_{\mm} > 60~{\mathrm GeV}/c^2, 
\end{array}
\end{displaymath}

\noindent in agreement with the Standard Model predictions, computed with \mbox{GRC4F}: 

\begin{displaymath}
\begin{array}{cr}
 \sigma
 = 0.112~{\mathrm pb} &  15 < m_{\mm} < 60~{\mathrm GeV}/c^2, \\
\sigma 
 = 0.033~{\mathrm pb} &  m_{\mm} > 60~{\mathrm GeV}/c^2. 
\end{array}
\end{displaymath}
%
%\noindent Due to low statistics the values of the cross-sections in
% the leptonic channel at $\sqrt{s} = 182.6$~GeV were not derived.

%DB \newpage
%         Modified on 04-06-1999 by dimartino
\section{Acknowledgements}
\vskip 3 mm
We are greatly indebted to our technical
collaborators, to the members of the CERN-SL Division for the excellent
performance of the LEP collider and to the funding agencies for their
support in building and operating the DELPHI detector.\\
We acknowledge in particular the support of \\
Austrian Federal Ministry of Education, Science and Culture,
GZ 616.364/2-III/2a/98, \\
FNRS--FWO, Flanders Institute to encourage scientific and technological
research in the industry (IWT), Belgium,  \\
FINEP, CNPq, CAPES, FUJB and FAPERJ, Brazil, \\
Czech Ministry of Industry and Trade, GA CR 202/96/0450 and GA AVCR A1010521,\\
Commission of the European Communities (DG XII), \\
Direction des Sciences de la Mati$\grave{\mbox{\rm e}}$re, CEA, France, \\
Bundesministerium f$\ddot{\mbox{\rm u}}$r Bildung, Wissenschaft, Forschung
und Technologie, Germany,\\
General Secretariat for Research and Technology, Greece, \\
National Science Foundation (NWO) and Foundation for Research on Matter (FOM),
The Netherlands, \\
Norwegian Research Council,  \\
State Committee for Scientific Research, Poland, 2P03B06015, 2P03B11116 and
SPUB/P03/DZ3/99, \\
JNICT--Junta Nacional de Investiga\c{c}\~{a}o Cient\'{\i}fica
e Tecnol$\acute{\mbox{\rm o}}$gica, Portugal, \\
Vedecka grantova agentura MS SR, Slovakia, Nr. 95/5195/134, \\
Ministry of Science and Technology of the Republic of Slovenia, \\
CICYT, Spain, AEN96--1661 and AEN96-1681,  \\
The Swedish Natural Science Research Council,      \\
Particle Physics and Astronomy Research Council, UK, \\
Department of Energy, USA, DE--FG02--94ER40817. \\

\newpage
%%%%%%%%%%%%%%%%%%%%%%%%%%%%%%%%%%%%%%%%%%%%%%%%%%%%%%%%%%%%%%%%%%%%%%%%%%
%BIBLIOGRAPHY
%%%%%%%%%%%%%%%%%%%%%%%%%%%%%%%%%%%%%%%%%%%%%%%%%%%%%%%%%%%%%%%%%%%%%%%%%%

%=========================================================================%

\newpage

\end{document}